# A modified Schottky model for graphene-semiconductor (3D/2D) contact: A combined theoretical and experimental study

Shi-Jun Liang[1*], Wei Hu[2], A. Di Bartolomeo[3], Shaffique Adam[4] and Lay Kee Ang[1]

[1]SUTD-MIT International Design Centre, Singapore University of Technology and Design, Singapore, [2]Lawrence Berkeley National Laboratory, California, USA, [3]Physics Department E.R. Caianiello, University of Salerno and CNR-SPIN Salerno, Fisciano, Italy, [4]Centre for Advanced 2D materials, National University of Singapore and Yale-NUS college, Singapore,

*Email: shijun_liang@mymail.sutd.edu.sg, Tel: +65 86221838

*Abstract*— In this paper we carry out a theoretical and experimental study of the nature of graphene/semiconductor Schottky contact. We present a simple and parameter-free carrier transport model of graphene/semiconductor Schottky contact derived from quantum statistical theory, which is validated by the quantum Landauer theory and first-principle calculations. The proposed model can well explain experimental results for samples of different types of graphene/semiconductor Schottky contact.

## INTRODUCTION

Due to a zero bandgap, graphene (**Gr**) can be regarded as a semimetal. So far **Gr**-semiconductor (3D or 2D) Schottky contact has been widely used in transistors/barristors, solar cells, sensors and other electronic/optoelectronic devices [1, 2]. Recent experiments have pointed out that the measured Richardson constant for **Gr**/Si (A=0.02 A/cm$^2$/K$^2$) [3] and **Gr**/MoSe$_2$ (A=0.024 A/cm$^2$/K$^2$) [4] Schottky contact is much smaller than A=112 (60) A/cm$^2$/K$^2$ given by traditional Schottky diode equation. However, researchers still cling to traditional Schottky diode equation out of convenience and are hesitant to look beyond the traditional thermionic emission model for a more consistent picture with graphene's unique properties. The following questions: *(1) does the traditional Schottky diode equation hold for **Gr**-semiconductor (3D or 2D) Schottky junction? (2) How does the Schottky barrier height (SBH) form at the interface? (3) What are the advantages of **Gr**/2D semiconductor junction over **Gr**/3D semiconductor junction, in terms of efficient charge transport?* still have not been satisfactorily answered.

Here we present a theoretical and experimental study of **Gr**-semiconductor contact for different semiconductors (different materials and dimensionality). Based on experiments and first-principle calculations, we propose a modified Schottky diode equation valid for any **Gr**/semiconductor contact, which is in excellent agreement with experimental results.

## MODEL

### A. Schottky diode equation for Gr/semiconductor contact

The traditional Schottky diode equation contains the terms describing the carrier's mass in the material.

$$J = AT^2 \exp\left(-\frac{\phi_{Bn}}{k_B T}\right)\left[\exp\left(\frac{qV}{nk_B T}\right) - 1\right] \quad (1)$$

$$A = \frac{4\pi q m k_B^2}{h^3}$$

Direct application of this equation to graphene will definitely lead to contradiction with massless carrier property. Besides, the ultrafast Fermi velocity is not manifested in the traditional Schottky diode equation. Furthermore, the failure of conventional Schottky diode equation becomes clear when applied to the description of experimental I-V characteristics of **Gr**-semiconductor Schottky junctions.

To get the supply function for electron emission from graphene plane, we consider that the electrons move in the graphene plane direction (behaving as massless Fermions), but are confined in the quantum well in the perpendicular direction. According to recently developed thermionic emission theory for graphene [5], the current density from semiconductor to graphene is given by

$$J_{S-Gr} = \int_{q\phi_B}^{\infty} q\, dn(E_z) \quad (2)$$

$$dn(E_z) = \frac{1}{\pi \hbar^3 v_f^2} \int_{E_z}^{\infty} \frac{(E - E_z)dE}{\exp\left(\frac{E - E_F}{k_B T}\right) + 1}$$

Where E and E$_z$ are the total energy and normal energy component of electrons, respectively, $v_f$ is the Fermi velocity, $\phi_{Bn}$ is the SBH. Assuming that the energy tail of the Fermi-Dirac distribution is important, Eq. (2) can be reduced down to

$$J_{S-Gr} = A^* T^3 \exp\left(-\frac{\phi_B - E_F}{k_B T}\right) \quad (3)$$

$$A^* = \frac{q k_B^3}{\pi \hbar^3 v_f^2}$$

In the thermal equilibrium, the current density flowing from graphene to semiconductor must be equal to the one from semiconductor to graphene. Thus for a **Gr**-semiconductor Schottky junction under a bias, the traditional Schottky diode equation for a metal semiconductor contact (given by Eq. (1)) becomes

$$J = A^* T^3 \exp\left(-\frac{\phi_B - E_F(V)}{k_B T}\right)\left[\exp\left(\frac{qV}{nk_B T}\right) - 1\right] \quad (4)$$

The novel and most relevant feature of Eq. (4) is the inclusion of ultrafast Fermi velocity and massless carrier properties as well as bias-tunable Fermi level in the graphene. Remarkably, using the basic physical constant, $A^* = 0.01158$ A/cm$^2$/K$^3$ is given, a value close to the mentioned experimental data for **Gr**/Si or **Gr**/MoSe$_2$ Schottky contact.

## B. Experimental validation of model in A

Fig. 1 shows the temperature behavior of the I-V characteristics of a **Gr**/Si junction fabricated by transferring Cu-CVD monolayer graphene on an array of Si-nanotips etched on a n-type Si-wafer as detailed in ref. [6]. The measurements were taken in dark and at

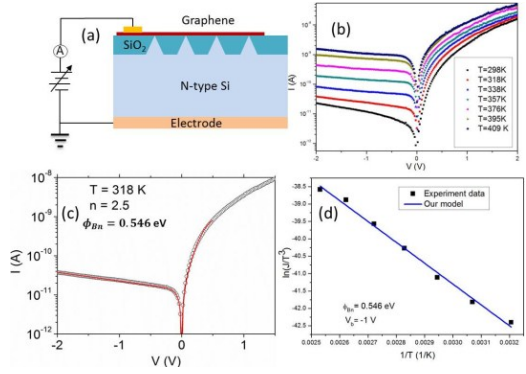

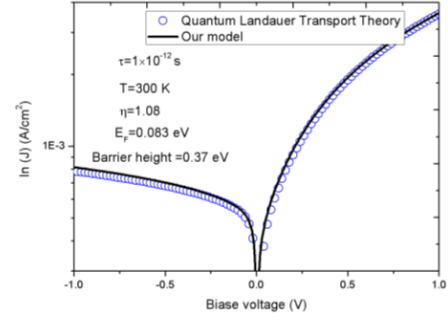

Fig. 3 Current density (ln(J)) for graphene/Si Schottky junction as a function of bias voltage using our model (solid line) and quantum Landauer transport theory (blue circle) [2].

Fig. 1 (a) Setup of the **Gr**/n-Si tips Schottky junction device (b) I-V characteristics of device shown in (a) at different temperatures. (c) I-V curve (red solid line) obtained using Eq. (4) with linear dependence of $E_F$ on V shows an excellent agreement with measured curves (open circle symbol) at T=318 K. The I-V curve at high forward bias is dominated by series resistance and high injection and is not included in the fit. (d) Temperature dependence of the reverse current at reverse bias of -1 V (ln(I/T$^3$) versus 1/T). The extracted reverse current data (square black symbol) is fitted with Eq. (4) (blue line).

atmospheric pressure.

Clearly, the model developed in this paper is able to reproduce the measured curves of ln(J/T$^3$) versus 1/T in Fig. 1. Furthermore Eq. (4), with a linear dependence of $E_F$ on V, can well describe the measure I-V characteristics at 318 K as shown in Fig. 1c. Despite the excellent agreement with the experimental data of graphene/Si-tips junction of Fig. 1, to check the robustness of the model, we applied it to other sets of experimental data relative to flat, larger-area **Gr**/Si, **Gr**/MoS2, **Gr**/GaAs and **Gr**/GaN contact, as shown in Fig. 2. All of these experimental data shows good agreement with our model. These plots strongly support the argument that Eq. (4) well describes the fundamental transport mechanisms across a **Gr**/semiconductor Schottky junction.

Furthermore, we validate our model by comparing it to the predications of *Quantum Landauer Transport Theory [2]*. Remarkably, we find that they are in excellent agreement as demonstrated in **Fig. 3**.

## C. Origin of inhomogeneous SBH of graphene/semiconductor contact

Recently, experiments have confirmed that the Schottky barrier at the interface of **Gr**/Si, **Gr**/GaAs and **Gr**/Ge contact is spatially inhomogeneous. Ripple topography in the graphene layer and interfacial disorder are suggested to be two key factors [7, 8] determining the formation of spatially inhomogeneous SBH. However, we believe that the origin of inhomogeneous SBH is rather electron-hole puddles distribution, caused by the charge- impurities randomly distributed on the surface of semiconductor. The presence of electron-hole puddles induces the fluctuations in charge density across graphene sample. As a result, the Fermi level is no longer constant across the whole graphene layer, but fluctuates from site to site. We consider that the barrier distribution is continuous at the interface between graphene and semiconductor. Moreover, we assume that the spatial distribution graphene's Fermi level follows a Gaussian distribution $P(E_F) = \frac{1}{\sqrt{2\pi}\delta_P}\exp\left(-\frac{(E_F-\overline{E_F})^2}{2\delta_P^2}\right)$, with standard deviation $\delta_P$ around a mean $\overline{E_F}$ value. The physical model is presented in Fig. 4 (left panel). By incorporating Fermi level fluctuation model into graphene's thermionic emission theory, the Schottky diode equation describing **Gr**/semiconductor junction with inhomogeneous SBH at the interface is given by

$$J_{S-Gr} = \int_{-\infty}^{\infty} P(E_F)dE_F \frac{q}{\pi\hbar^3 v_f^2} \int_{q\phi_B}^{\infty} dE_z \int_{E_z}^{\infty} \frac{(E-E_z)dE}{\exp\left(\frac{E-E_F}{k_BT}\right)+1}$$

$$J = A^*T^3 \exp\left(-\frac{\overline{\phi_B}-\frac{\delta_P^2}{2k_BT}}{k_BT}\right)\left[\exp\left(\frac{qV}{nk_BT}\right)-1\right] \quad (5)$$

Using Eq. (5), we can reproduce the experimental data for **Gr**/Si ($\delta_P$=135 meV), **Gr**/GaAs ($\delta_P$=98 meV) and **Gr**/Ge ($\delta_P$=85 meV) contact system, as can be seen from **Fig. 4**. Furthermore, we can estimate the impurity charge concentration $\sigma_s$ distributed on the surface of three different semiconductors. The mean SBH for each contact is also given. The excellent agreement of our model with experimental indicates that our model is applicable to charge transport on every Gr/semiconductor Schottky contact.

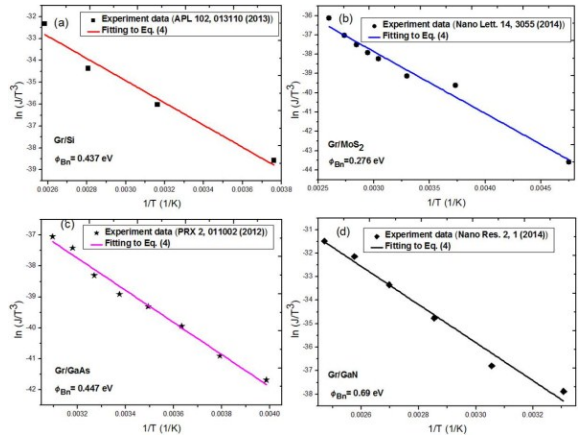

Fig. 2 ln(J/T$^3$) versus 1/T for experimental data (symbols) and theoretical results (solid lines) (a) **Gr**/Si Schottky diod. (b) **Gr**/MoS2 contact. (c) **Gr**/GaAs contact. (d) **Gr**/GaN contact.

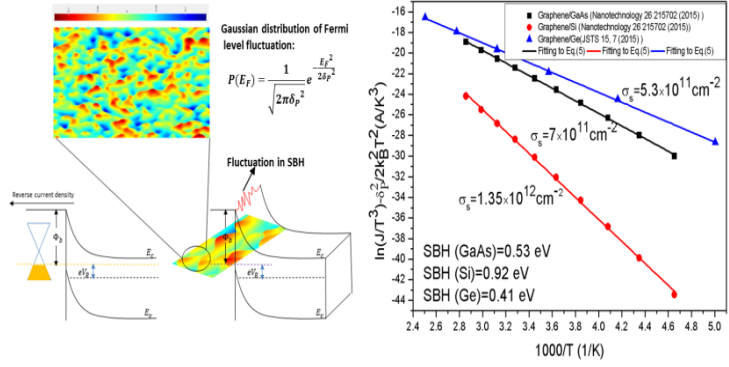
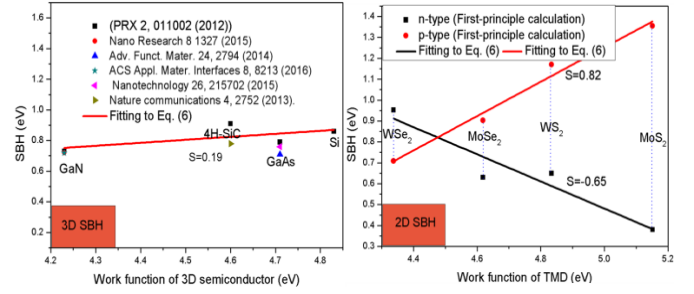

Fig.4. (Left panel) Two-dimensional band diagram of spatially inhomogeneous **Gr**/semiconductor Schottky contact. Spatial variation (Gaussian distribution) in Fermi level across the graphene leads to different SBH. (Right panel) Experiment data (symbols) for **Gr**/Si, **Gr**/GaAs and **Gr**/Ge contact and theoretical results (solid lines).

Fig. 5. Correlation between SBH and work function of semiconductor. (a) SBH vs work function of 3D semiconductor in **Gr**/semiconductor (GaN, 4H-SiC, GaAs and Si) Schottky contact, (b) SBH vs work function of 2D semiconductor in **Gr**/(n- and p-type)semiconductor ($WSe_2$, $MoSe_2$, $WS_2$ and $MoS_2$) Schottky contact.

D. **Formation mechanism of SBH at the interface of Gr/semiconductor contact**

As in the traditional metal/semiconductor Schottky contact, the carrier current flowing across graphene/semiconductor interface is exponentially dependent on SBH, which determines the current-voltage characteristics. When graphene is brought into contact with a 3D semiconductor, the energy band alignment at the interface represents an interesting challenge, due to the mismatch in dimensionality. Furthermore, the formational mechanism for **Gr**/3D-semiconductor and **Gr**/2D-semiconductor can be definitely distinctive. For real applications in electronics/optoelectronics, a comprehensive and deep understanding of the formation of SBH can offer a precise control of I-V characteristics. Unfortunately, we are still far from a complete understanding of the formation of SBH at the graphene and semiconductor interface. Here we employ first-principle calculations and available experimental data to distinguish the formation mechanism of SBH for **Gr**/3D-semiconductor and **Gr**/2D-semiconductor contact. The Schottky barrier height on a n- or p- type semiconductor can be expressed as

$$\phi_{Bn/p} = E_0 + S(W_S - E_0) \qquad (6)$$

Where $\phi_{Bn/p}$ is SBH, $E_0$ is the reference energy, S is the degree of correlation between SBH and work function of the semiconductor, and $W_S$ is the work function of the semiconductor. In **Fig. 5a** and **b** we examine the correlation between SBH and work function for **Gr**/semiconductor contact in 2D and 3D case. Data suggests that the SBH for **Gr**/3D-semiconductor is less correlated to work function, implying that SBH is dependent on atomic structure of graphene and 3D semiconductor interface. In this case, new chemical bonds are formed at the interface, or atomic orbits belonging to graphene and semiconductors are hybridized or overlapped to some degree. The weak dependence of SBH on the work function of 3D semiconductor is an indication that Fermi level pinning might take place in 3D case. In contrast, the SBH is strongly correlated to work function of the 2D semiconductor (e.g. transition metal dichalcogenide) in **Gr**/2D-semiconductor contact (see **Fig. 5b**), which suggests an interface free from Fermi pinning effect. In this latter case, SBH can be uniquely identified by the work function of 2D semiconductor for given graphene. In **Fig. 6**, **7**, **8**, we calculated DOS vs energy, differential charge density vs z position and effective potential vs z position based on first-principle calculation.

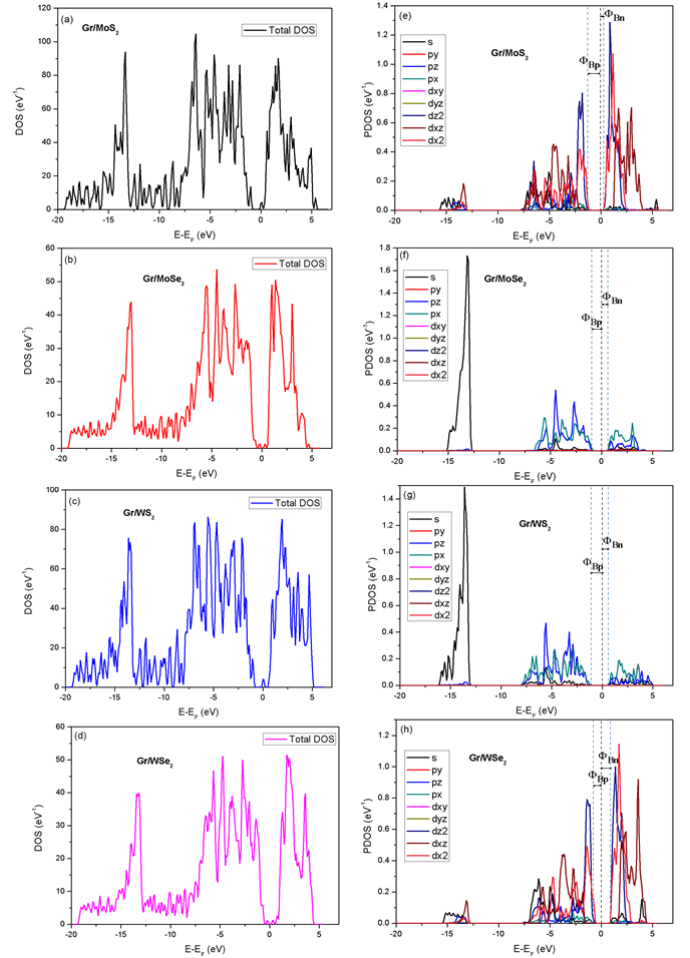

Fig. 6. Density of state and PDOS vs energy with respect to Fermi level. (a) and (e) Gr/$MoS_2$, (b) and (f) Gr/$MoSe_2$, (c) and (g) Gr/$WS_2$, (d) and (h) Gr/$WSe_2$

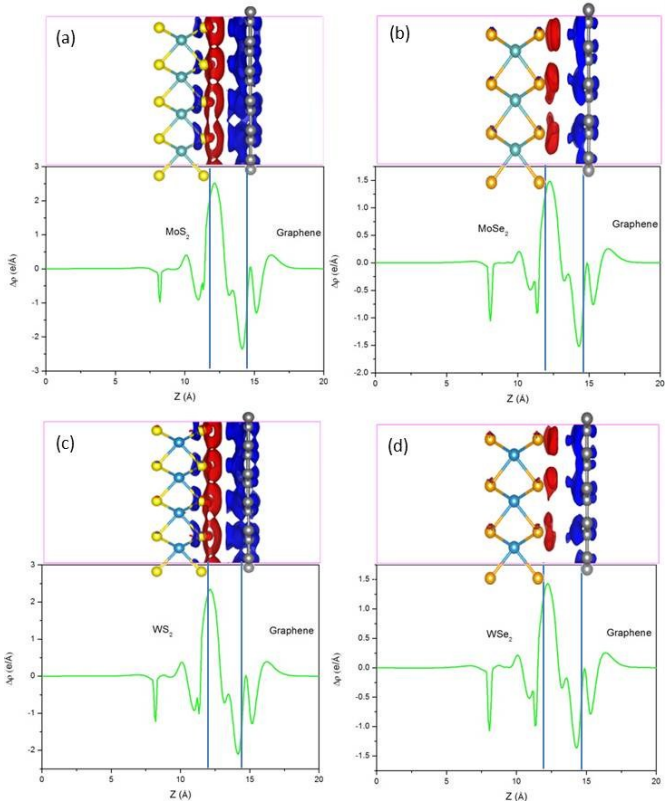

Fig. 7. Differential charge density of Gr/TMD contact at equilibrium interfacial distance. (a) Gr/$MoS_2$, (b)Gr/$MoSe_2$, (c) Gr/$WS_2$, (d) Gr/$WSe_2$.

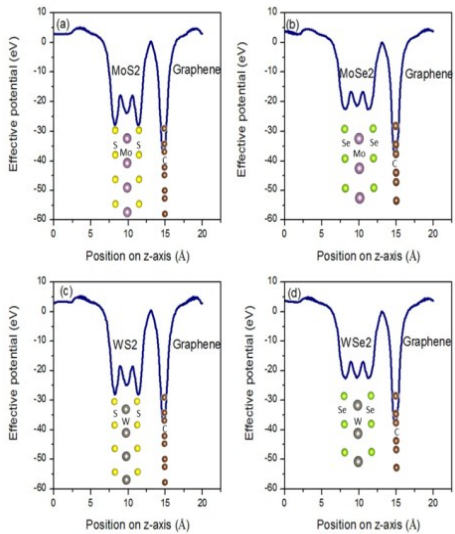

Fig. 8. Plot of effective potential versus z position for Gr/TMD contact. (a) Gr/$MoS_2$, (b)Gr/$MoSe_2$, (c) Gr/$WS_2$, (d) Gr/$WSe_2$.

## Conclusion

In this paper, we have used a combined experimental and theoretical approach to study the contact between graphene and (3D/2D) semiconductor. Some new perspectives are presented to improve the understanding of the nature of **Gr**/semiconductor contact as follows:

- Traditional Schottky diode equation for metal/semiconductor contact is no longer satisfactory for **Gr**/semiconductor Schottky contact. A new equation, consistent with experimental data, has been proposed to account for the unique characteristics of graphene and semiconductor Schottky contact.

- A physical model is proposed to account for the origin of spatially inhomogeneous SBH at the interface of **Gr**/semiconductor contact. The proposed model is in excellent agreement with available experimental data.

- First-principle calculations and experiment data are employed to elucidate the formation mechanism of SBH for **Gr**/3D-semiconductor and **Gr**/2D-semiconductor contact.

- First-principle calculations have been used to develop a comprehensive understanding of the nature of electronic interface between graphene and 2D semiconductors (e.g. transition metal dichalcogenide).

### Acknowledgment

This work was supported by Singapore MOE T2 Grant (T2MOE1401). L. K. Ang acknowledges the support of AFOAR AOARD grant (14-2110).